\documentclass[aps,prd,reprint,twocolumn,superscriptaddress,showpacs,notitlepage]{revtex4-1}

\usepackage{graphicx}
\usepackage{mathrsfs}
\usepackage{bm}
\usepackage{amsmath}
\usepackage{dcolumn}
\usepackage{epstopdf}
\usepackage{dsfont}
\usepackage{amssymb}
\usepackage{tabularx}
\usepackage{array}
\usepackage{float}
\usepackage{color}
\usepackage{epstopdf}
\usepackage{mathrsfs}
\usepackage[colorlinks, linkcolor=blue,anchorcolor=blue,citecolor=blue,urlcolor=blue]{hyperref}
\usepackage[T1]{fontenc}

\usepackage{hyperref}
\hypersetup{hidelinks,
	colorlinks=true,
	allcolors=blue,
	pdfstartview=Fit,
	breaklinks=true}

\begin{document}
\title{M\textbf{\textit{O}}enes family materials with  Dirac nodal loop, strong light-harvesting ability, long carrier lifetime and conduction-band valley spin splitting}

\author{Luo Yan}
\email[]{yanluo@usc.edu.cn}
\affiliation{School of Mathematics and Physics, University of South China, Hengyang 421001, China}

\author{Junchi Liu}
\affiliation{School of Physics, University of Electronic Science and Technology of China, Chengdu 610054, China}

\author{Yu-Feng Ding}
\affiliation{School of Mathematics and Physics, University of South China, Hengyang 421001, China}

\author{Jiafang Wu}
\affiliation{College of Physics and Engineering, Chengdu Normal University, Chengdu, 611130, China}

\author{Bao-Tian Wang}
\affiliation{Institute of High Energy Physics, Chinese Academy of Science (CAS), Beijing 100049, China}

\author{Hao Gao}
\affiliation{Fritz-Haber-Institut der Max-Planck-Gesellschaft, Faradayweg 4-6, 14195 Berlin, Germany}

\author{Liujiang Zhou}
\email[]{liujiang86@gmail.com}
\affiliation{School of Physics, University of Electronic Science and Technology of China, Chengdu 610054, China}

\begin{abstract}
M\textbf{\textit{O}}enes, as emerging MXenes-like materials, also have wide structural spaces and various chemical and physical properties. Using first-principles and high-throughput calculations, we have built an online library (\url{https://moenes.online}) for M\textbf{\textit{O}}enes family materials from basic summaries, mechanical, phonon and electron aspects, based on their structural diversities from 2 stoichiometric ratios, 11 early-transition metals, 4 typical functional groups and 4 oxygen group elements. Compared to MXenes, the main advantage of M\textbf{\textit{O}}enes at present is that we have discovered 14 direct semiconductors, which greatly increases the number of direct semiconductors and the range of band gap values in the MXenes family.  Among them, 1T-Ti$_{2}$\textit{\textbf{O}}F$_{2}$ (\textbf{\textit{O}}=O, S, Se) reveal tunable semiconducting features and strong light-harvesting ability ranging from the ultraviolet to the near-infrared region. Besides, 2H- and 1T-Y$_{2}$TeO$_{2}$ have a long carrier lifetime of 2.38 and 1.24 ns, originating from their spatially distinguished VBM and CBM states and long dephasing times. In addition, 2H-Zr$_{2}$O(O)$_{2}$ shows spin-valley coupling phenomena, and the valley spin splitting is apparent and robust in its conduction band ($\sim$85 meV). Therefore, M\textbf{\textit{O}}enes have a wealth of physical properties, not limited to those reported here, and future studies of these emerging M\textbf{\textit{O}}enes are appealing.
\end{abstract}
\maketitle

\section{Introduction}
The structural space of the two-dimensional (2D) transition metal carbides, nitrides and carbonitrides (MXenes) has a wide range of control possibilities, from abundant early-transition metals, stoichiometric ratios, surface functional groups and solid solution formations\cite{naguib2011two,halim2016synthesis,naguib2021ten}. Still now, more than 40 types of MXenes with distinct stoichiometry have been successfully synthesized in experiments, and the theoretically predicted number of MXenes is even more unpredictable in view of the formation of its solid solution\cite{naguib2021ten}. In addition, the structural diversity of MXenes gives rise to their variable and tailorable physical and chemical properties.  As a result, MXenes have a wide range of applications ranging from energy storage and harvesting to catalysis, water purification and desalination, electromagnetic interference shielding, communications, optics, electronics, plasmonics, sensors, actuators, composites, and biomedicine\cite{jiang2020two, vahidmohammadi2021world}. However, only Sc$_{2}$C(OH)$_{2}$\cite{lee2015achieving}, Cr$_{2}$TiC$_{2}$(OH)$_{2}$\cite{yang2016tunable} and Y$_{2}$C(OH)$_{2}$\cite{Wang2019} MXenes are direct semiconductors and their band gap values are mainly in the range of 0.71$-$0.84 eV (HSE06), which severely limits MXenes' applications in the semiconductor field.

\begin{figure*}[t!]
	\centering
	\includegraphics[width=0.8\linewidth]{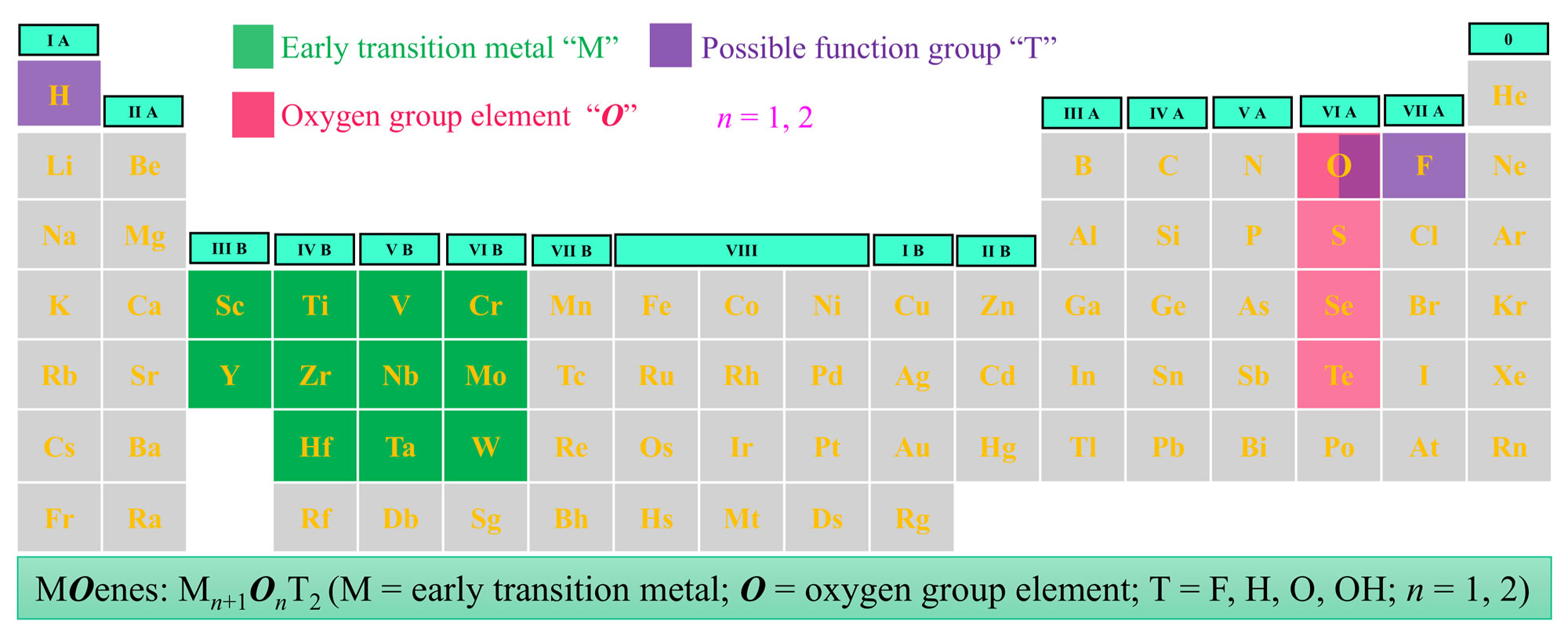}
	\caption{The compositions of M\textit{\textbf{O}}enes in the periodic table. The green, purple and pink frame represent early transition metal, possible function group (-O, -H, -OH, -F) and oxygen group element (O, S, Se, Te), respectively.}
\end{figure*}

Prior to this work, we have extended traditional MXenes to 2D early transition metal-based oxides, M\textbf{\textit{O}}enes, relying on the regulation of the "X" position in MXenes. Correspondingly, the MXenes-like M\textit{\textbf{O}}enes have a general formula of  M$_{n+1}$\textbf{\textit{O}}$_{n}$T$_{x}$,  where M indicates an early-transition metal,where \textit{\textbf{O}} is the oxygen group element (O, S, Se and Te), and T$_{x}$  is the surface functional group, such as hydrogen, hydroxyl, oxygen or fluorine (Fig. 1).  In fact, the layer 1T-Ti$_{2}$O phases \cite{fan2019structure,wang2020controllable}, oxycarbide MXenes \cite{michalowski2022oxycarbide} and Zr$_{2}$Se(B$_{1-x}$Se$_{x}$) (\textit{x}= 0$-$0.97) chalcogenide MAX phases strongly support the experimental feasibility of this emerging  M\textit{\textbf{O}}enes family. Moreover, bare Ti$_{2}$O M\textit{\textbf{O}}enes have promise for electrides, superconductor and anode materials in lithium and sodium batteries \cite{yan2020single}. Importantly, Ti$_{2}$OX$_{2}$ (X = F, Cl) M\textit{\textbf{O}}enes and their Janus phases are direct semiconductors with a band gap of 0.58$-$1.18 eV at the hybrid functional (HSE06) level, together with quantum phase transitions, long carrier lifetime, strong light-harvesting ability \cite{yan2022two,yan2023direct} and superior thermoelectric performance \cite{wan2024high,wan2024enhancement}. Terminated 2H-Ti$_{2}$O M\textit{\textbf{O}}enes and their Janus structures also show potentials in piezoelectric transducers and actuators \cite{qiu2024new}. Besides, 2H-Ti$_{n+1}$O$_{n}$ (\textit{n} = 1$-$3) shows a slightly weakened superconductivity with increasing atomic thickness, 2H-Ti$_{3}$O$_{2}$F$_{2}$/Ti$_{4}$O$_{3}$F$_{2}$ exhibits quantum spin Hall effects at room temperature \cite{yan2024thickness}. Oxidized 2H-M$_{2}$O (M= Ti and Zr) M\textit{\textbf{O}}enes show strong linear and non-linear optical response in the infrared range \cite{huang2024prediction}. Furthermore, when the early-transition metal turns to V, the oxygen-terminated V$_{2}$O M\textit{\textbf{O}}enes are perfect electrocatalysts for the hydrogen evolution reaction \cite{xie2024two}.  But, not just limited to these, the compositions of M\textit{\textbf{O}}enes also have highly adjustable space as shown in Fig. 1, and the role of the different "M" and "\textit{\textbf{O}}" sites in the M\textit{\textbf{O}}enes is still not clear. Besides, research on M\textit{\textbf{O}}enes has only just begun, to make the M\textit{\textbf{O}}enes family we are interested in better known, their fundamental properties, including stability, mechanical properties and electronic structures, are very effective for future research. Therefore, an online library of materials will be a practical way to make getting to know M\textit{\textbf{O}}enes easier. Significantly, M\textbf{\textit{O}}enes with direct band gaps are excellent complements to the MXenes. Therefore, the development of the excellent direct semiconductors of M\textit{\textbf{O}}enes beyond what we have explored will further expand the application prospects of the MXenes family in the semiconductor field.

In this work, 2H- and 1T-M$_{\textit{n}+1}$\textbf{\textit{O}}$_{\textit{n}}$T$_{2}$ (\textit{n} = 1, 2) M\textit{\textbf{O}}enes are fully scrutinized, including 11 early-transition metals, 4 typical functional groups and 4 oxygen group elements (Fig. 1). Using first-principles and high-throughput calculations, the M\textit{\textbf{O}}enes are studied and summarized from basic summaries, mechanical properties, phonon spectra and electronic structures. Furthermore, the online materials library for M\textit{\textbf{O}}enes has also been built to quickly understand the properties of the desired M\textit{\textbf{O}}enes. Of the total of 880 M$_{\textit{n}+1}$\textbf{\textit{O}}$_{\textit{n}}$T$_{2}$ (\textit{n} = 1, 2) M\textit{\textbf{O}}enes studied here, 464  are dynamically stable and show semimetallic, metallic, semiconducting and (topological) insulating features.  The effects of the "M" and "O" sites on the M\textbf{\textit{O}}enes have also been studied and compared. In particular, 13 direct semiconductors are newly found beyond Ti$_{2}$OX$_{2}$ (X = F, Cl) \cite{yan2022two} and Ti$_{2}$OFCl M\textit{\textbf{O}}enes \cite{yan2023direct}, which is a perfect compensation for the defects of MXenes. The direct semiconductors are rare and their bandgap values are concentrated. In this regard, our focus here is on these direct semiconductors with exotic properties, and other aspects merit further research. In particular, 1T-Ti$_{2}$SF$_{2}$ and 1T-Ti$_{2}$SeF$_{2}$  have strong infrared and visible light absorption properties by considering electron-hole (e-h) interactions, which can be used in infrared detection and solar energy fields. In addition, the weak nonadiabatic coupling (NAC) induced by spatially distinct valence band maximum (VBM) and the conduction band minimum (CBM) states and the long dephasing time prolong the carrier lifetimes for 2H and 1T-Y$_{2}$TeO$_{2}$ M\textbf{\textit{O}}enes, reaching the nanosecond level. Furthermore, the coupled spin and valley physics in 2H-ZrO(O)$_{2}$ M\textbf{\textit{O}}enes provides another platform to investigate spintronic and valleytronic properties. Unlike MXenes, M\textbf{\textit{O}}enes exhibit rich semiconducting features and are suitable for future researches in electronics, optoelectronics, photovoltaics, valleytronics, etc.

\section{Results and discussion}
\begin{figure}[t!]
	\begin{center}
		\includegraphics[width=1\linewidth]{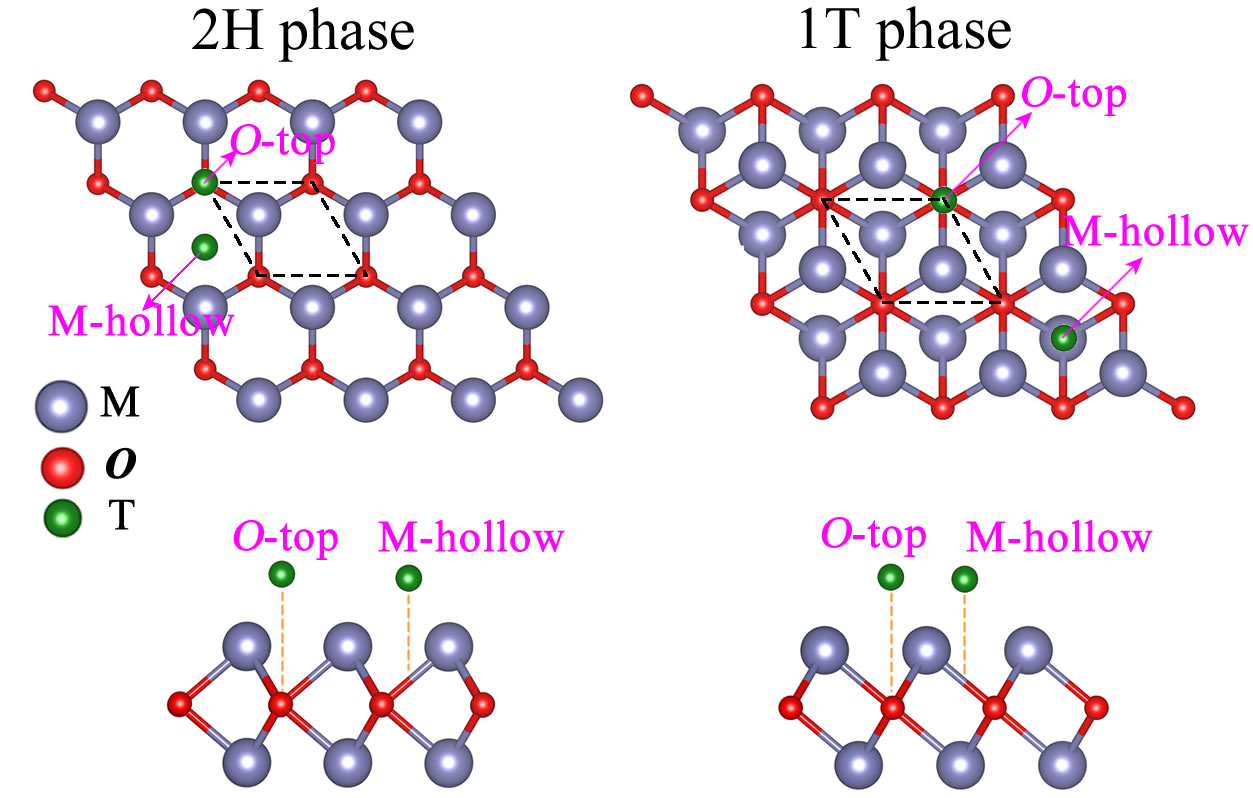}
	\end{center}
	\caption{Top view of (a) 2H- and (b) 1T-M$_{\textit{n}+1}$\textbf{\textit{O}}$_{\textit{n}}$T$_{2}$ (\textit{n} = 1, 2; T = F, H, OH and O) M\textbf{\textit{O}}enes. Here, the side view of M\textbf{\textit{O}}enes is depicted from (c) 2H- and (d) 1T-M$_{2}$\textbf{\textit{O}}T$_{2}$. The two possible functionalization sites on either side of the surface are \textbf{\textit{O}}-top (on the top of the \textbf{\textit{O}} sites ) and M-hollow (on the hollow site of the hexatomic ring of M and \textbf{\textit{O}} sites). The black line means the primitive cell.}
\end{figure}

\begin{figure*}[htp!]
	\centering
	\includegraphics[width=0.8\linewidth]{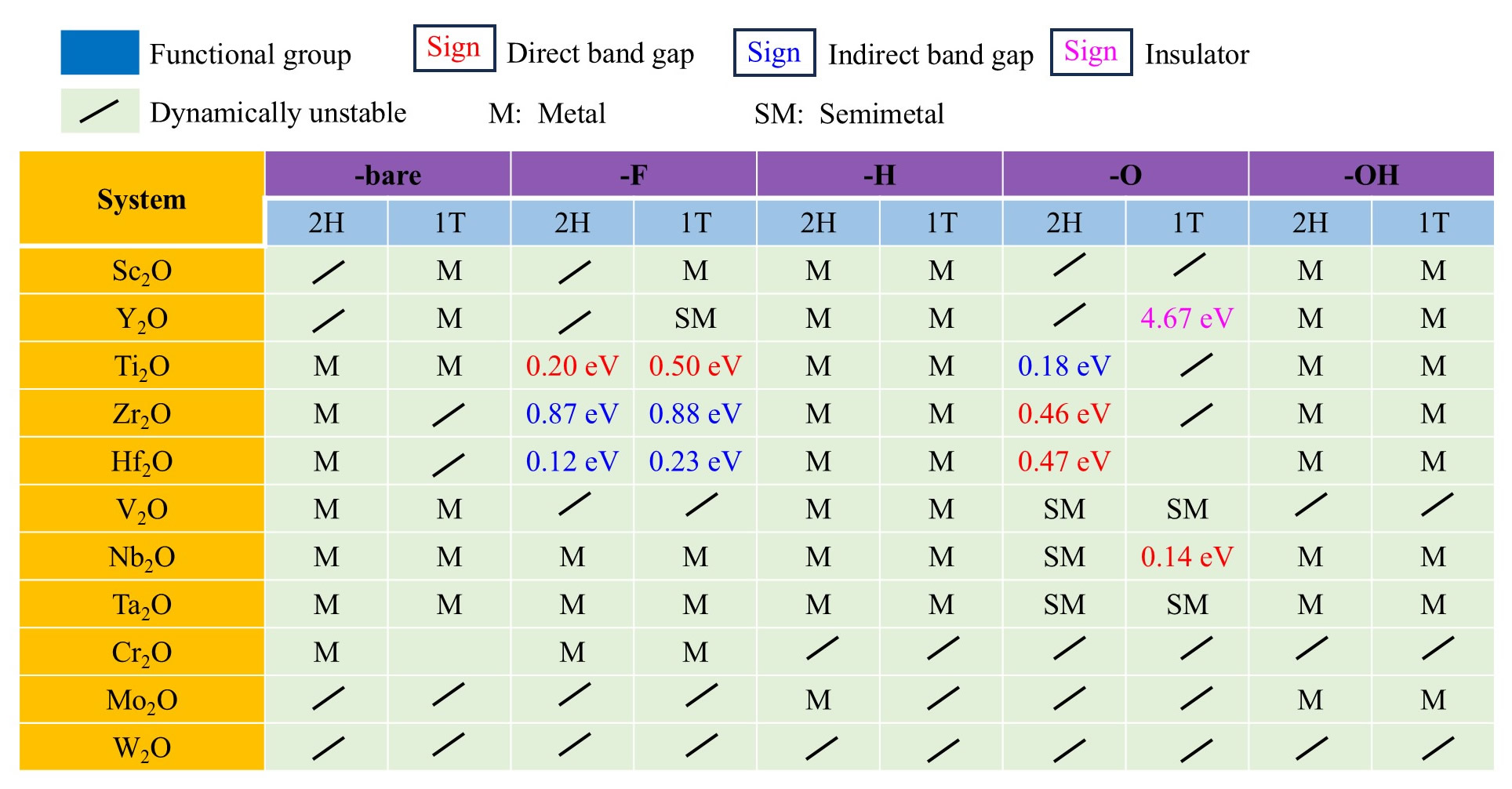}
	\caption{Summarised information for M$_{2}$OT$_{2}$ M\textbf{\textit{O}}enes, including stability and electronic structural properties distinguished by direct/indirect semiconducting, insulating, metallic and semimetallic features. The red, blue, and purple values represent the band gap values of  direct semiconductors, indirect semiconductor, and insulator, respectively.}
\end{figure*}

\subsection{Crystal structures of M\textbf{\textit{O}}enes}
The crystal structures of M$_{\textit{n}+1}$\textbf{\textit{O}}$_{\textit{n}}$T$_{2}$  M\textbf{\textit{O}}enes are shown in Fig. 2. In general, M\textbf{\textit{O}}enes have two phases: 2H trigonal prismatic and 1T octahedral phases. As shown in Fig. 2a, the 2H phase consists of edge-sharing trigonal prisms belonging to the space group $P\overline6m2$ (No. 184). On the other hand, 1T phase has a space group of $P\overline3m1$ (No. 164), and is composed of [M\textit{\textbf{O}}$_{6}$] octahedra. Due the unsaturated orbitals of outer M metals, bare M\textbf{\textit{O}}enes are apt to be terminated by functional groups. In this respect, the most two possible functionalisation sites are considered, as exhibited in Fig. 2. The most energetically stable crystal structures of M$_{\textit{n}+1}$\textbf{\textit{O}}$_{\textit{n}}$T$_{2}$ M\textbf{\textit{O}}enes can be determined by comparing their energies. In the following, we fully studied the bare M$_{\textit{n}+1}$\textbf{\textit{O}}$_{\textit{n}}$ M\textbf{\textit{O}}enes and terminated M$_{\textit{n}+1}$\textbf{\textit{O}}$_{\textit{n}}$T$_{2}$ M\textbf{\textit{O}}enes from dynamical, mechanical and electronic features by high-throughput calculations.

\subsection{Online materials library for M\textbf{\textit{O}}enes}
To give a quick understanding for this emerging M\textbf{\textit{O}}enes family, we firstly have built an online materials library (\url{http://moenes.online}). The workflow for investigating  M\textbf{\textit{O}}enes is shown in Fig. S1. Based on the most energetically stable  M$_{\textit{n}+1}$\textbf{\textit{O}}$_{\textit{n}}$ and M$_{\textit{n}+1}$\textbf{\textit{O}}$_{\textit{n}}$T$_{2}$ M\textbf{\textit{O}}enes, we firstly confirm the ground sates of M\textbf{\textit{O}}enes by comparing energies between nonmagnetic (NM) and  ferromagnetic (FM) spin configurations. Phonon spectra of  M\textbf{\textit{O}}enes are next used to confirm their dynamical stability. Besides, their mechanical stability are determined by Born criteria for hexagonal crystal, namely, $C_{11}$ > |$C_{12}$| and $C_{66}$ = ($C_{11}$$-$$C_{12}$)/2 > 0 \cite{mouhat2014necessary}. When M\textbf{\textit{O}}enes have reliable dynamical and mechanical stability, the electronic properties of M\textbf{\textit{O}}enes are explored, and then their band characters can be determined by the locations of the VBM and CBM).

In brief, the structures of 880 M\textbf{\textit{O}}enes have been presented in this online materials library, from basic summaries to mechanical properties, phonon spectra and band structures. One can quickly search for the characteristics of the desired M\textbf{\textit{O}}enes by selecting stoichiometric ratios (\textit{n}), early-transition metals (M), oxygen group elements (\textit{\textbf{O}}) and functional groups (T) step by step. Taking Ti$_{2}$SF$_{2}$ M\textbf{\textit{O}}enes as illustrations, the search results are shown in Fig. S2. In basic summaries part, we can obtain the optimized lattice constant, favorable absorption site, the most favourable dynamical stability, magnetic moments of early-transition metal, locations of VBM and CBM, band character, band gap in semiconductors/insulators and work functions, etc. In mechanical properties part, the elastic constants, mechanical stability, Young's modulus (\textit{Y}) and possion's ratio ($\nu$) are available. In addition, we can acquire their phonon dispersions within the phonon spectra, and projected band structures (Pbands) are provided as well.

The stability, exotic electronic features including semimetals, direct/indirect semiconductors and insulators are also summarized. The summaries for M$_{2}$OT$_{2}$  M\textbf{\textit{O}}enes are shown in Fig. 3. There are 70 M$_{2}$OT$_{2}$  M\textbf{\textit{O}}enes are dynamically stable. Obviously, most of them have rich metallic properties, along with several semimetals, such as 1T-Y$_{2}$OF$_{2}$, 2H-M$_{2}$O(O)$_{2}$ (M=V, Nb, Ta) and 1T-M$_{2}$O(O)$_{2}$ (M=V, Ta), in which the Weyl fermion locates at the Fermi level. Besides, there are 10 semiconductors and one insulators. Among them, 2H/1T-Ti$_{2}$OF$_{2}$, 2H-M$_{2}$O(O)$_{2}$ (M = Zr, Hf) and 1T-Nb$_{2}$O(O)$_{2}$ are direct semiconductor with a band gap of 0.14$-$0.50 eV(PBE).  2H/1T-M$_{2}$OF$_{2}$ (M = Zr, Hf) and 2H-Ti$_{2}$O(O)$_{2}$ are indirect semiconductors. Besides, 1T-Y$_{2}$O(O)$_{2}$ is an insulator with a band gap of 4.67 eV (PBE), and futher corrected to  6.20 eV (HSE06). It is worth emphasizing that the summaries for M$_{2}$ST$_{2}$, M$_{2}$SeT$_{2}$, M$_{2}$TeT$_{2}$, M$_{3}$O$_{2}$T$_{2}$, M$_{3}$S$_{2}$T$_{2}$, M$_{3}$Se$_{2}$T$_{2}$ and  M$_{3}$Te$_{2}$T$_{2}$ M\textbf{\textit{O}}enes are available in Supporting Informations. As listed in Table 1 and shown in Fig. S11, there are 14 direct semiconductors in the M\textbf{\textit{O}}enes family, and their band gap values are 0.32$-$1.28 eV (HSE06), reflecting their promising applications in semiconducting fields. Furthermore, they have a small Y (< 260 N/m) and isotropic $\nu$, meaning their soft properties, and thus can applied in flexible devices \cite{lee2008measurement}.

\begin{table}[htp]
	\setlength{\tabcolsep}{8pt}
	\caption{The bandgaps of direct semiconductors, evaluated by PBE ($E_{g}$ in eV) and HSE06 ($E_{g}^{hse}$ in eV), along with their Y (N/m) and $\nu$.}
	\begin{center}
		\begin{tabular}{ccccccccccccccccccc}
			\hline
			Systems&$E_{g}$&$E_{g}^{hse}$&Y&$\nu$&\\
			\hline
			2H-Ti$_{2}$OF$_{2}$&0.20&0.82&230.80&0.19\\
			1T-Ti$_{2}$OF$_{2}$&0.50&1.18&229.90&0.18\\
			2H-Zr$_{2}$O(O)$_{2}$&0.46&1.28&149.80&0.21\\
			2H-Hf$_{2}$O(O)$_{2}$&0.47&1.24&139.79&0.26\\
			1T-Nb$_{2}$O(O)$_{2}$&0.14&0.90&251.96&0.37\\
			1T-Ti$_{2}$SF$_{2}$&0.15&0.67&170.60&0.17\\
			1T-Zr$_{2}$SF$_{2}$&0.35&0.78&168.94&0.19\\
			2H-Hf$_{2}$SF$_{2}$&0.11&0.38&179.32&0.25\\
			1T-Ti$_{2}$SeF$_{2}$&0.33&0.97&164.48&0.15\\
			2H-Zr$_{2}$SeF$_{2}$&0.17&0.42&141.74&0.24\\
			1T-Zr$_{2}$SeF$_{2}$&0.70&1.23&166.88&0.16\\
			2H-Y$_{2}$TeO$_{2}$&0.84&1.27&137.40&0.37\\
			1T-Y$_{2}$TeO$_{2}$&0.79&1.18&142.28&0.36\\
			2H-Hf$_{3}$Se$_{2}$F$_{2}$&0.15&0.32&233.57&0.22\\
			\hline
		\end{tabular}
	\end{center}
\end{table}

As presented above, the bare M\textbf{\textit{O}}enes are metals and usually lack a good stability, but they can transform into dynamically stable phases when their outer orbitals of M are saturated by functional groups. On the other hand, the electronic traits of M\textbf{\textit{O}}enes are strongly related to the M. In short, the majority of fluorinated and hydroxidized Ti-, Zr-, Hf-based M\textbf{\textit{O}}enes show semiconducting features, and other fluorinated and hydroxidized M\textbf{\textit{O}}enes are mostly metallic. In addition, oxidized Sc- and Y-based M$_{2}$\textit{\textbf{O}} M\textbf{\textit{O}}enes have insulating and semiconducting  characters, but oxidized V-, Nb-, Ta-based M$_{2}$\textit{\textbf{O}} M\textbf{\textit{O}}enes possess semimetal and metallic features. All in all, due to the structural diversity of M\textbf{\textit{O}}enes, it brings the diversity of its electronic structures, including highly conductive metallic, semimetallic, semiconducting, (topological) insulating, which can bring various applications for M\textbf{\textit{O}}enes as traditional MXenes.

\subsection{The effect of "M" site on M\textbf{\textit{O}}enes}
\begin{figure*}[t!]
	\centering
	\includegraphics[width=0.8\linewidth]{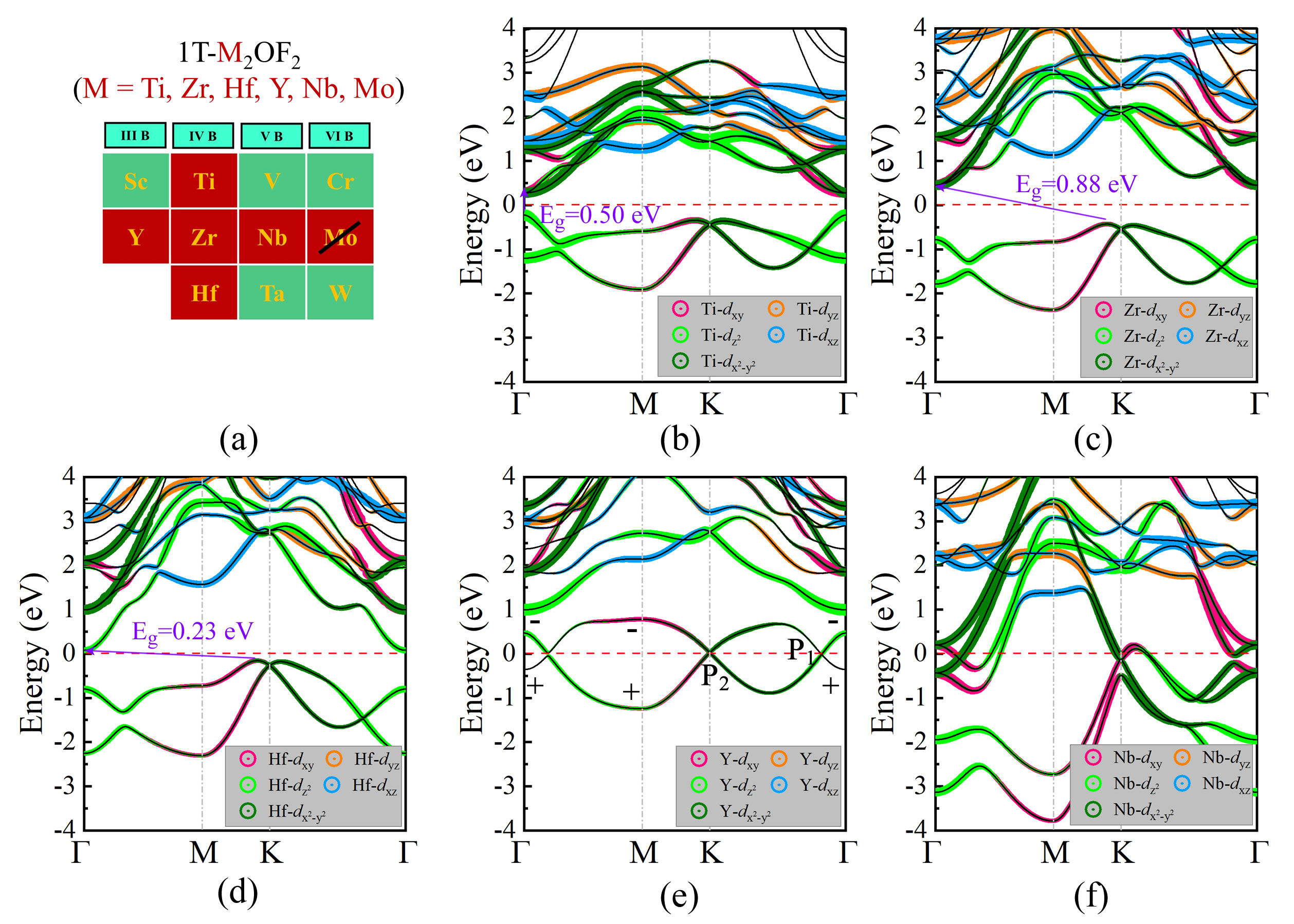}
	\caption{(a) Selected early transition metals in the periodic table. Pands at PBE level for (b) 1T-Ti$_{2}$OF$_{2}$, (c) 1T-Zr$_{2}$OF$_{2}$, (d) 1T-Hf$_{2}$OF$_{2}$, (e) 1T-Y$_{2}$OF$_{2}$ and (f) 1T-Nb$_{2}$OF$_{2}$, weighted by their \textit{d} orbitals. Signs "+" and "-" in  (e) represents the symmetries of parity. The Fermi level is set to 0 eV.}
\end{figure*}

Since early transition metal elements bring rich electronic structures in M\textbf{\textit{O}}enes, we have selected group IVB metal elements (Ti, Zr, Hr) and fifth-row transition metal elements (Y, Zr, Nb, Mo) under the chemical formula 1T-M$_{2}$OF$_{2}$ to fully characterize the influence of "M" sites in  M\textbf{\textit{O}}enes (Fig. 4a). In all fluorinated 1T-M$_{2}$O M\textbf{\textit{O}}enes, because of the lower electronegativity of the transition metals than the O atom and the attached fluorinated functional groups, the transition metals become positively charged by donating their electrons to both O atoms and the fluorinated functional groups. Firstly, Ti, Zr, and Hf are in the same IVB group in the periodic table with the same number of electrons in their outer shells ([Ar]3d$^{2}$4s$^{2}$, [Kr]4d$^{2}$5s$^{2}$, and [Xe]4f$^{14}$5d$^{2}$6s$^{2}$). As expect, fluorinated 1T-Zr$_{2}$O and Hf$_{2}$O exhibit similar semiconducting properties to those of 1T-Ti$_{2}$OF$_{2}$\cite{yan2022two}.  Moreover, their energy states around the Fermi energy are mainly made of the M-\textit{d}$_{z^{2}}$ and M-\textit{d}$_{x^{2}-y^{2}}$ (M = Ti, Zr and Hf) orbitals. As shown in Fig. 4b, 1T-Ti$_{2}$OF$_{2}$ is a direct semiconductor together with VBM and CBM located at the $\Gamma$ point. However, when M changes from Ti, to Zr, and even to Hf, the M-\textit{d}$_{z^{2}}$ orbitals move down, while the M-\textit{d}$_{x^{2}-y^{2}}$  orbitals move up. As a result, the M-\textit{d}$_{x^{2}-y^{2}}$ at the K point is higher than the M-\textit{d}$_{z^{2}}$ at the $\Gamma$ point in. Thus,  1T-Zr$_{2}$OF$_{2}$ and 1T-Hf$_{2}$OF$_{2}$ become indirect semiconductors (Fig. 4c and 4d).

Around the Zr metal element, the fifth-row transition metal elements (Y, Zr, Nb) are further systematically scrutinized to draw the roles of the different number of electrons in their outer shells ([Kr]4d$^{1}$5s$^{2}$, [Kr]4d$^{2}$5s$^{2}$, [Kr]4d$^{4}$5s$^{2}$). Due to the dynamical instability in 1T-Mo$_{2}$OF$_{2}$, we don't consider it further. Clearly shown in Pbands, as the number of electrons in Y, Zr and Nb outer shells progressively increases, the Fermi energy gradually shifts upwards and the \textit{d}$_{x^{2}-y^{2}}$ orbitals near the Fermi energy move down largely. As Zr changes to Nb metal, the increase of Fermi energy and the movement of Nb-\textit{d}$_{x^{2}-y^{2}}$ orbitals result in the metal characters for 1T-Nb$_{2}$OF$_{2}$ (Fig. 4f). When the Zr turns to Y, the Fermi energy shifts downwards and cuts directly through the middle of the most highest valance bands in 1T-Zr$_{2}$OF$_{2}$, leading to a semimetal feature for 1T-Y$_{2}$OF$_{2}$ (Fig. 4e).  Clearly, two salient features can be observed in 1T-Y$_{2}$OF$_{2}$ without SOC (Figs. 4e and S12a).  Firstly, 1T-Y$_{2}$OF$_{2}$ has two linear band crossings in the Fermi level (P$_{1}$ and P$_{2}$). Secondly, these linear band crossings are observed for an electron and hole band near the Fermi level. A careful scan of the band structure shows that the intersection P$_{1}$ is not isolated, but is actually part of an approximate nodal loop centred on the $\Gamma$ point, as indicated in Figs. S12c and S12e. However, , the approximate nodal ring is not complete, and the exact crossing points are located only on the high symmetry path, such as P$_{1}$. Remarkably, the crossing points (labeled N$_{1}$ and N$_{2}$) still persist when SOC is turned on (Fig. S12b), which is similar to that without SOC due to the weak SOC effect in 1T-Y$_{2}$OF$_{2}$. The difference is that each band, when considering SOC, is doubly degenerate. Moreover, a complete nodal ring (labeled L$_{1}$) exists around the $\Gamma$ point at this time, as shown by the calculated energy gap between the electron-like and hole-like bands in Fig. S12f. Since each band is doubly degenerate, the nodal ring is fourfold degenerate Dirac nodal ring, as illustrated in Fig. S12d. In addition, the Dirac nodal ring of this system lies exactly on the Fermi surface and there are no non-topological bands near the Fermi energy, reflecting the properties of an ideal 2D Dirac nodal loop state.

\subsection{The effect of "\textbf{\textit{O}}" site on M\textbf{\textit{O}}enes}

\begin{figure*}[t!]
	\centering
	\includegraphics[width=\linewidth]{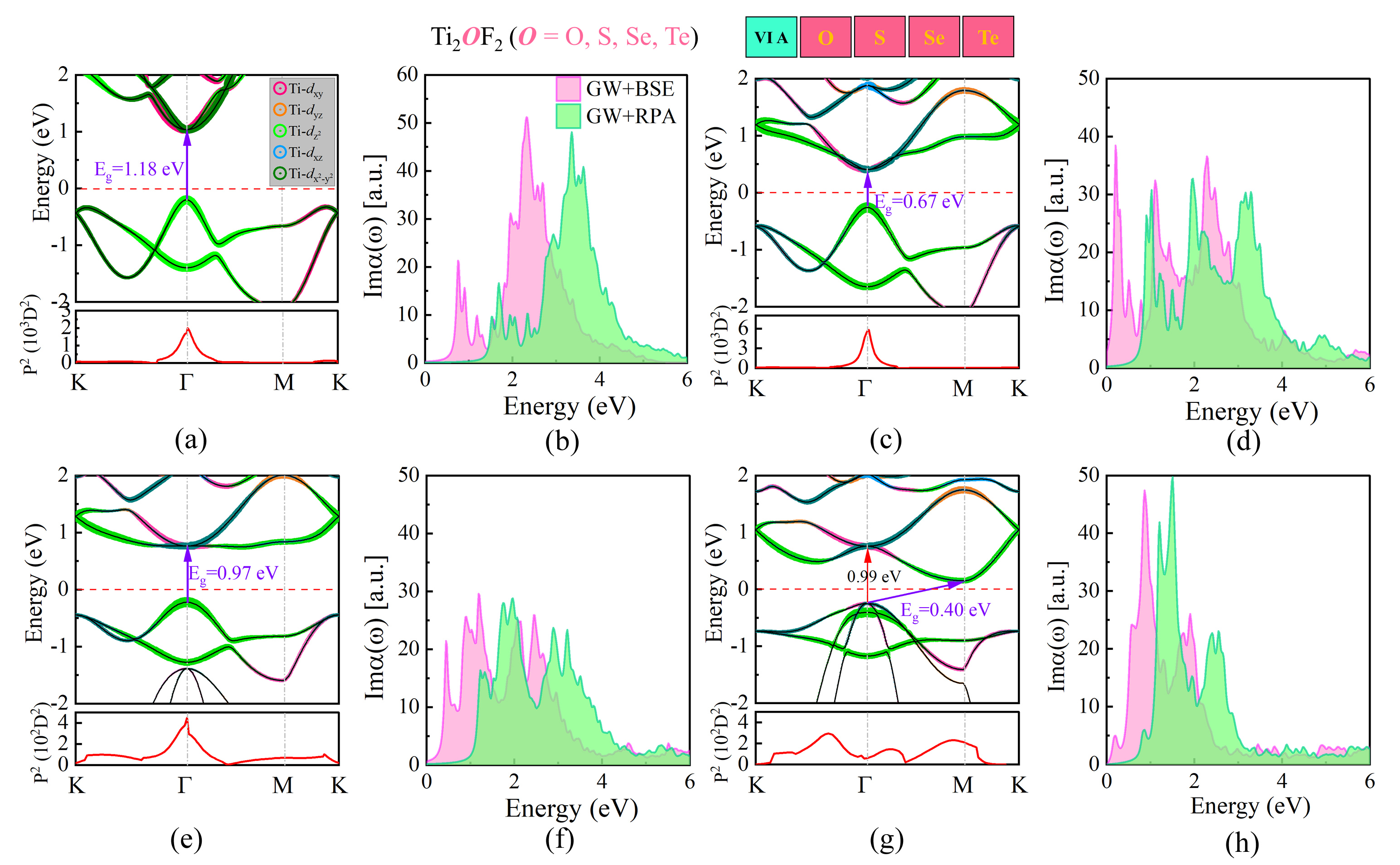}
	\caption{Pbands (upper plane) along with \textit{P}$^{2}$ (lower plane)  for (a) 1T-Ti$_{2}$OF$_{2}$, (c) 1T-Ti$_{2}$SF$_{2}$, (e) 1T-Ti$_{2}$SeF$_{2}$ and (g) 1T-Ti$_{2}$TeF$_{2}$, calculated at the HSE06 level.  Imaginary part of the macroscopic polarizability of (b) 1T-Ti$_{2}$OF$_{2}$, (d) 1T-Ti$_{2}$SF$_{2}$, (f) 1T-Ti$_{2}$SeF$_{2}$ and (h) 1T-Ti$_{2}$TeF$_{2}$ calculated on the basis of GW+RPA and GW+BSE methods.}
\end{figure*}

In order to draw the influence of oxygen group elements, the properties of Ti$_{2}$\textbf{\textit{O}}F$_{2}$ (O = O, S, Se, Te) are systematically scrutinized and compared. We first focused on the 1T phase. Firstly, 1T-Ti$_{2}$SF$_{2}$, 1T-Ti$_{2}$SeF$_{2}$ and 1T-Ti$_{2}$TeF$_{2}$ can maintain structural integrity at 900, 600 and 300 K, respectively (Fig. S13), lower than that of 1T-Ti$_{2}$OF$_{2}$ (1200 K)\cite{yan2022two}. Clearly, as the oxygen group elements change from O to S, Se and Te, the thermal stability is weakened due to the increase in chemical bonds. (Table II). Similar to 1T-Ti$_{2}$OF$_{2}$ shown in Fig. 5a \cite{yan2022two}, 1T-Ti$_{2}$SF$_{2}$ and 1T-Ti$_{2}$SeF$_{2}$ are direct semiconductors with VBM and CBM located at the $\Gamma$ point, and have a band gap of 0.67 and 0.97 eV (HSE06), respectively (Figs. 5c and 5e). The sum of square of the transition dipole moment (\textit{P}$^{2}$) also confirm the allowed electronic transition between VBM and CBM at the $\Gamma$ point. Their VBM states are composed of Ti-\textit{d}$_{z^{2}}$ orbitals mostly, but CBM states have hybrid orbitals from Ti-\textit{d}$_{xy}$, Ti-\textit{d}$_{z^{2}}$ and Ti-\textit{d}$_{x^{2}-y^{2}}$. The same as 1T-Ti$_{2}$OF$_{2}$ monolayer \cite{yan2022two}, three low-energy bands are involved in the band edges at $\Gamma$ point, including one valence band and two degenerate conduction bands composed by one heavy electron-band and one light electron-band. Besides, 1T-Ti$_{2}$TeF$_{2}$ monolayer has an indirect band gap of 0.40 eV (HSE06), which is also reflected by the results of \textit{P}$^{2}$  (Fig. 5g).  As expected, their band gaps at the $\Gamma$ point first decrease and then increase when the oxygen group elements change from O to S, Se and then to Te. This is due to the decrease in electronegativity of the oxygen group elements, which weakens the bond strength in the 1T-Ti$_{2}$\textit{\textbf{O}}F$_{2}$ (\textbf{\textit{O}}= O, S, Se, Te) M\textbf{\textit{O}}enes (Table II). The clearer Pbands are shown in Fig. S14. As the oxygen group elements change from O to S, the light electron-band contributed by Ti-\textit{d}$_{z^{2}}$ orbitals moves down and gradually becomes a heavy electron-band, leading to a decrease of the band gap firstly. When the oxygen group elements change from S to Se, this band continuously moves down and the orders of the original two degenerate conduction bands are changed, bringing an increase of the band gap. In this progress, the band dispersions of two degenerate conduction bands and valence band around the $\Gamma$ point have been significantly weighted, leading to a larger effective mass and a smaller carrier mobility. Finally, 1T-Ti$_{2}$TeF$_{2}$ monolayer becomes an indirect semiconductor with VBM and CBM located at the $\Gamma$ and M point, respectively.

As the oxygen group elements change progressively, the optical absorption behaviour of 1T-Ti$_{2}$\textbf{\textit{O}}F$_{2}$ (O = O, S, Se, Te) are further compared within a GW method simulated with the random phase approximation (GW+RPA) (without e-h interactions) and the Bethe-Salpeter equation (GW+BSE) (with e-h interactions). The imaginary parts of the frequency-dependent 2D polarizability for 1T-Ti$_{2}$SF$_{2}$, 1T-Ti$_{2}$SeF$_{2}$ and 1T-Ti$_{2}$TeF$_{2}$ are shown in Figs. 5d, 5f and 5h, respectively, and compared with that for 1T-Ti$_{2}$OF$_{2}$ (Fig. 5a)\cite{yan2022two}. Here, the optical band gap $E_{g}^{o}$ is defined as the first dominated peak corresponding to the first optically allowed (bright) exciton. The $E_{g}^{o}$ values for 1T-Ti$_{2}$SF$_{2}$, 1T-Ti$_{2}$SeF$_{2}$ and 1T-Ti$_{2}$TeF$_{2}$ are 0.21 and 0.45 eV, respectively. In term of the spectral edge obtained at the GW+RPA level, the quasi-particle energy gap ($G_{0}W_{0}$ gap) are extracted to 0.74 (1.05) eV for  1T-Ti$_{2}$S(Se)F$_{2}$. Here, GW correction to the PBE band gap is about 0.59 (0.70) eV. Then, the exciton binding energy $E_{b}$ values for 1T-Ti$_{2}$SF$_{2}$ and 1T-Ti$_{2}$SeF$_{2}$ are evaluated to 0.53 and 0.59 eV, respectively. As listed in Table II, the numerical values of the $E_{g}^{o}$ and $G_{0}W_{0}$ gap show the same trend as the band gap values. Generally, the larger the band gap, the weaker the screening, resulting in a higher $E_{b}$. Therefore, the $E_{b}$ values of 1T-Ti$_{2}$\textit{\textbf{O}}F$_{2}$ (\textbf{\textit{O}} = O, S, Se) decrease firstly, and then increase, on going from O to Se.  Similar to 1-Ti$_{2}$OF$_{2}$ monolayer\cite{yan2022two}, there are several significant peaks in the infrared and visible light regions, making 1T-Ti$_{2}$\textit{\textbf{O}}F$_{2}$ (\textbf{\textit{O}}=S, Se, Te) promising for infrared detectors and solar cells with strong light absorption ability in the desired spectral range.

\begin{figure*}[htp!]
	\centering
	\includegraphics[width=1\linewidth]{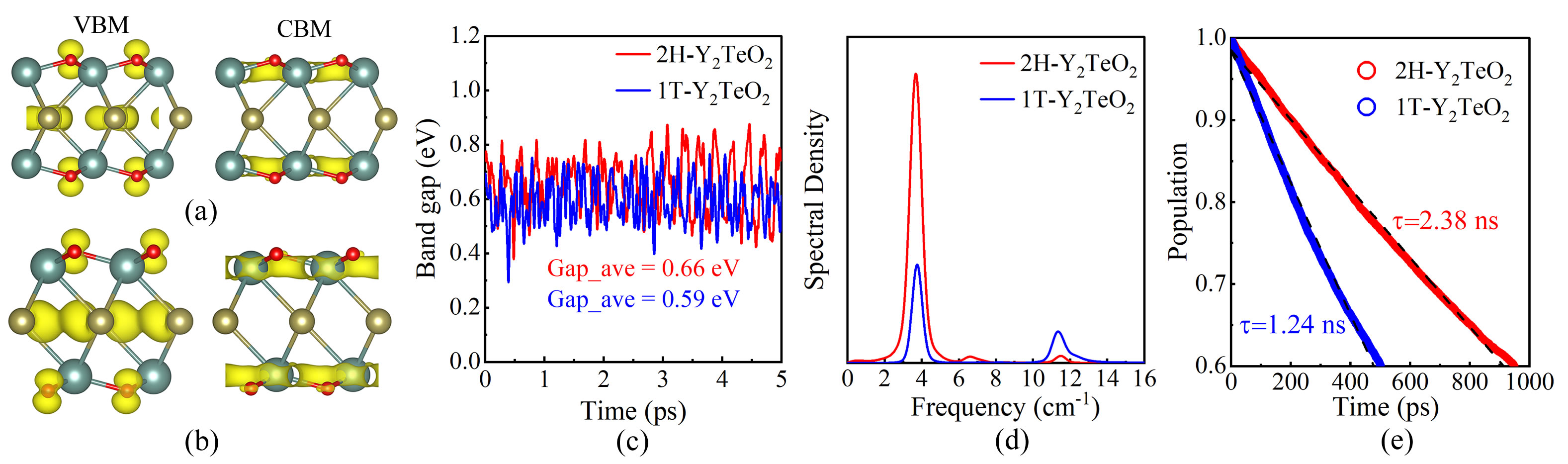}
	\caption{Band decomposed charge density of VBM and CBM at the $\Gamma$ point for (a) 2H- and (b) 1T-Y$_{2}$TeO$_{2}$. (c) The variations of band gaps during the 5ps' AIMD simulations. (d) Spectral densities and (e) excited-electron population decay.} 
\end{figure*}

On the other hand, as the oxygen group elements gradually change from O to Te, the 2H-Ti$_{2}$\textit{\textbf{O}}F$_{2}$ (\textbf{\textit{O}} = O, S, Se, Te) changes from a direct semiconductor to a Weyl semimetal, and finally to an indirect semiconductor (Fig. S15) at the PBE level, undergoing a similar evolution of the Ti-\textit{d}$_{z^{2}}$ orbitals discussed above. When spin-orbit coupling (SOC) is not considered, the Weyl fermions in 2H-Ti$_{2}$SF$_{2}$ and 2H-Ti$_{2}$SeF$_{2}$ are purely located at the $\Gamma$ point at the Fermi level (Figs. S16a and 16d). At this time, the VBM and CBM touch quadraticall and form a two-component 2D double Weyl fermion with an $E^{\prime}$ irreducible representation of the $D_{3h}$ point group. When SOC is considered, these degenerate Weyl fermions contributed by Ti-\textit{d}$_{xy}$ and Ti-\textit{d}$_{x^{2}-y^{2}}$ orbitals are split, resulting in a SOC gap of $\sim$19 and 9 meV (PBE) for 2H-Ti$_{2}$SF$_{2}$ and 2H-Ti$_{2}$SeF$_{2}$, respectively (Figs. S16b and S16e). Moreover, in Figs. S16c and 16f, the edge states depicted 2H-Ti$_{2}$SF$_{2}$ and 2H-Ti$_{2}$SeF$_{2}$  connect the bulk valence and conduction bands. Besides, the $Z_{2}$ invariant values are determined to be 1 via Wannier charge center calculations. Thus, 2H-Ti$_{2}$SF$_{2}$ and 2H-Ti$_{2}$SeF$_{2}$ are determined to be 2D topological insulators, whose nontrivial gaps are further corrected to $\sim$24 and 55 meV (HSE06). Although thses nontrivial gaps are smaller than that in Ti$_{3}$O$_{2}$F$_{2}$ and Ti$_{4}$O$_{3}$F$_{2}$ M\textbf{\textit{O}}enes, they are adequate for practical applications at room temperature \cite{yan2024thickness}.

\begin{table}[htp]
	\setlength{\tabcolsep}{6pt}
	\caption{Optical band gaps $E_{g}^{o}$ (eV), $G_{0}W_{0}$ gap (eV), exciton binding energy $E_{b}$ (eV), and chemical bond length (\AA) for 1T-Ti$_{2}$\textit{\textbf{O}}F$_{2}$ (\textbf{\textit{O}} = O, S, Se, Te) M\textbf{\textit{O}}enes. }
	\begin{center}
		\begin{tabular}{ccccccccccccccccccc}
			\hline
			Systems&$E_{g}^{o}$&$G_{0}W_{0}$ gap &$E_{b}$&Ti$-$$\textit{\textbf{O}}$&Ti$-$F\\
			\hline
			1T-Ti$_{2}$OF$_{2}$&0.76&1.37&0.61&2.15&2.09\\
			1T-Ti$_{2}$SF$_{2}$&0.21&0.74&0.53&2.47&2.15\\
			1T-Ti$_{2}$SeF$_{2}$&0.45&1.05&0.59&2.60&2.17\\
			1T-Ti$_{2}$TeF$_{2}$&0.19&0.64&0.45&2.79&2.19\\
			\hline
		\end{tabular}
	\end{center}
\end{table}

\subsection{Exotic direct semiconductors in M\textbf{\textit{O}}enes}
As summarized before, the most significant advantage identified thus far is that M\textbf{\textit{O}}enes markedly enhance the scope of direct semiconductors and light absorption within the MXenes family. Therefore, we put our attention on these direct semiconducting  M\textbf{\textit{O}}enes with fascinating properties in the following.

\subsubsection{Long carrier lifetime in 2H- and 1T-Y$_{2}$TeO$_{2}$ M\textbf{\textit{O}}enes}
Carrier lifetime is critical in optoelectronic devices, where nonradiative (NA) electron-hole (e-h) recombination is the dominant factor. In term of Fermi’s golden rule \cite{hyeon2009symmetric}, the NA e-h recombination rate is the e-h recombination rate is proportional to the square of the NAC constant, which is defined as \cite{zheng2019ab}:
\begin{align}
	d_{ij} = <\varphi_{j}|\frac{\partial}{\partial t}|\varphi_{k}> =\sum_{I} \dfrac{<\varphi_{j}|\nabla_{R_{I}}\mathcal{H} |\varphi_{k}>}{\varepsilon_{k} - \varepsilon_{j}}\dot{\textbf{R}_{I}},
\end{align}
where \textit{$\mathcal{H}$} is the Kohn-Sham Hamiltonian, $\varphi$$_{j/k}$ and $\varepsilon$$_{j/k}$ indicates the wave functions and eigenvalues for electronic states \textit{j}/\textit{k}, respectively. Here, the states \textit{j} and \textit{ k} are CBM and VBM, respectively. Obviously, smaller electron-phonon (e-ph) coupling elements (<$\varphi$$_{j}$|$\nabla$$_{R_{I}}$\textit{$\mathcal{H}$}|$\varphi$$_{k}$>) determined by the wave function overlap between VBM and CBM and the e-ph interactions, smaller nuclear velocity \textit{$\dot{\textbf{R}_{I}}$} that is roughly proportional to atomic mass \cite{guo2020tuning}, and larger energy gap ($\varepsilon$$_{k}$$-$$\varepsilon$$_{j}$) will lead to weaker NAC. 

\begin{figure*}[htp!]
	\centering
	\includegraphics[width=1\linewidth]{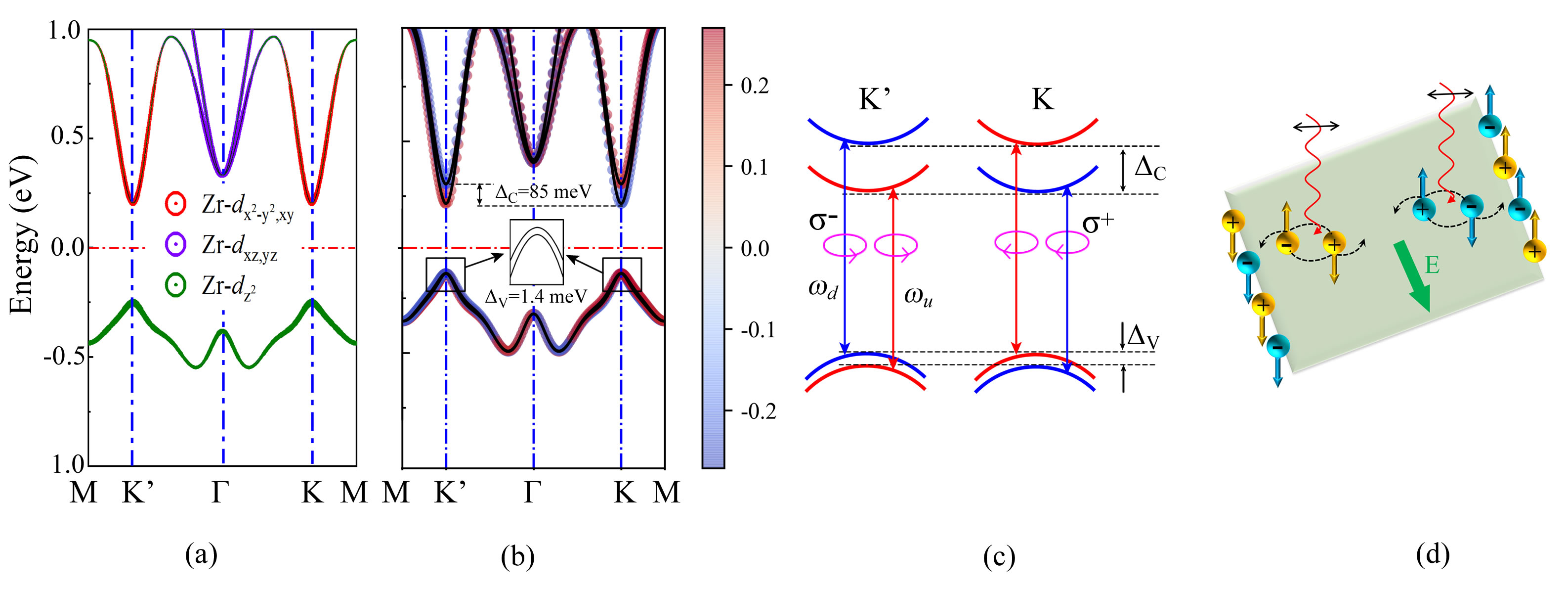}
	\caption{(a) Pbands and (c) spin projected band structure for 2H-Zr$_{2}$O(O)$_{2}$.  The blue color in (b) indicate the spin-down states, while red color denotes spin-up states (spin orientation is along out-of-plane direction). The Fermi level is set to 0 eV. (c) VSS in the VBM and the CBM at K/K' valleys, together with scheme of valley-dependent optical selection rules. Here, $\sigma$$^{+}$/$\sigma$$^{-}$ represents the right/left-hand polarized light, which couples to the band edge transition at the K/K' valley. (d) Spin and valley Hall effects under linearly polarized optical field with a frequency  $\omega$$_{u}$. Yellow and cyan balls indicate carriers in K' and K valleys, respectively, and the +/- symbol in the ball represents a hole/electron.}
\end{figure*}

Interestingly, the VB and CB band edge states of 2H- and 1T-Y$_{2}$TeO$_{2}$ M\textbf{\textit{O}}enes are occupied by different component (Figs. S11m and S11l), and as spatial charge density shown in Fig. 6a and 6b, the VBM states are mostly from Te atoms and O atoms, but the CBM states are fully contributed by Y atoms. Therefore, the spatially distinguished VBM and CBM states indicate a long carrier lifetime \cite{li2018time,yan2022two,yan2023theoretical,fu2024vertical}.  Firstly, 2H- and 1T-Y$_{2}$TeO$_{2}$ not only possess dynamical and mechanical stability, but also have a good stability at 300 K (Figs. S16a and S15b). Next, the NAC values for 2H- and 1T-Y$_{2}$TeO$_{2}$ are obtained to be 0.95 and 1.18 meV, respectively, which are greatly smaller than that in 2H-Ti$_{2}$OF$_{2}$ (2.46 meV) and 1T-Ti$_{2}$OF$_{2}$ (3.10 meV) \cite{yan2022two}. The fluctuations of band gap values (PBE) at 300 K are exhibited in Fig. 6c, and the averaged band gaps for 2H- and 1T-Y$_{2}$TeO$_{2}$ are 0.66 and 0.59 eV, respectively. Compared to the band gaps at 0 K (Table I), the band gaps decrease about 0.2 eV, due to the thermal distortions \cite{dou2021time}. In order to understand the phonon modes engaged in the e-h recombination and the relative e-ph interactions in Y$_{2}$TeO$_{2}$, the Fourier transforms of autocorrelation functions of the energy gap fluctuation along MD trajectory, known as spectral densities or influence spectra, is plotted in Fig. 6d. Clearly, a dominate peak appears at around 3.8 THz and a negligible peak locates at 11.7 THz, reflecting the low frequency regions contribute major rules in e-ph interactions. The vibrational modes around 3.8 THz at the $\Gamma$ point for 2H- and 1T-Y$_{2}$TeO$_{2}$ are visualized at Figs. S18a and S18b, respectively, where out-of-plane Y and O atom vibrations are participated in the e-h recombination. Moreover,  although the e-ph interaction for 2H-Ti$_{2}$OF$_{2}$ is stronger than that of 1T-Ti$_{2}$OF$_{2}$, 1T-Ti$_{2}$OF$_{2}$ has a larger band gap, resulting in a weaker NAC. All in all, the small NAC in Y$_{2}$TeO$_{2}$ M\textbf{\textit{O}}enes will greatly prolong the carrier lifetime.

By using Hefei-NAMD code \cite{zheng2019ab}, the carrier lifetimes of 2H- and 1T-Y$_{2}$TeO$_{2}$ are obtained. Generally, e-ph interactions are from inelastic and elastic scattering, where inelastic scattering transfers extra energy to involved nuclei, resulting in a NA recombination process; elastic scattering destroys the coherence in a quantum-mechanical superposition between VBM and CBM, leading to decoherence (also named as pure-dephasing) \cite{dou2021time}. Therefore, the decoherence has been considered into NAMD algorithm, and the decoherence time can be obtained by fitting the pure-dephasing functions with a Gaussian, exp[$-$0.5(\textit{t}/T)$^{2}$], plotted in Fig. S18c. The pure-dephasing time for 2H- and 1T-Y$_{2}$TeO$_{2}$ are calculated to 6.58 and 8.59 fs, respectively, which are smaller than that for 2H-Ti$_{2}$OF$_{2}$ (20.11 fs) and 1T-Ti$_{2}$OF$_{2}$ (11.26 fs) \cite{yan2022two}. When 2H-Y$_{2}$TeO$_{2}$ is compared to 1T-Y$_{2}$TeO$_{2}$, the transition energy with larger fluctuations (Fig. 7c) leads to a faster pure dephasing time, \textit{i.e.} a longer e-h recombination. Furthermore, the evolution of an excited electron state population is plotted to evaluate the carrier lifetime, as described in Fig. 6e. By fitting of a short-time linear expansion to the exponential decay,
\textit{f(t)}=exp($-$\textit{t}/$\tau$) $\approx$ 1$-$\textit{t}/$\tau$, the carrier lifetime for 2H- and 1T-Y$_{2}$TeO$_{2}$ are calculated to be 2.38 and 1.24 ns, respectively. Due to the weaker NAC and longer pure-dephasing time, their carrier lifetimes are much longer than that of 2H-Ti$_{2}$OF$_{2}$ (0.03 ns) and 1T-Ti$_{2}$OF$_{2}$ (0.37 ns) at the PBE level. Specially, the carrier lifetimes for 2H- and 1T-Y$_{2}$TeO$_{2}$ can reach the nanosecond scale, comparable to some 2D materials, such as pristine and doped BP (0.39$-$5.34 ns) \cite{guo2020tuning}, MoS$_{2}$ (0.16 ns), MoSe$_{2}$ (0.21 ns), MoSSe (0.24 ns) \cite{fu2024vertical}, and MoSTe (1.31 ns) \cite{jin2018prediction}. Therefore, long excited carrier lifetimes in Y$_{2}$TeO$_{2}$ M\textbf{\textit{O}}enes can facilitate the dissociation of excitons into free electron and hole carriers, favouring their fascinating performance in optoelectronic filed.

\subsubsection{Conduction-band valley spin splitting in 2H-Zr$_{2}$O(O)$_{2}$ M\textbf{\textit{O}}enes}
2H-Zr$_{2}$O(O)$_{2}$ and 2H-Hf$_{2}$O(O)$_{2}$ M\textbf{\textit{O}}enes are direct semiconductors together with the band energy extrema at the high symmetry K point ( Fig. S11c and 11d). The same symmetry as group IV and VI transition metal dichalcogenide monolayers (TMDs) \cite{xiao2012coupled,yuan2013zeeman,schaibley2016valleytronics}, the simultaneous existence of a direct band gap at the K point and a decoupled band edge would provide a platform for exploring valleytronic and spintronic features. However, 2H-Hf$_{2}$O(O)$_{2}$ are thermally unstable at room temperature (Fig. S17d). Therefore, in the following we will only concentrate on the 2H-Zr$_{2}$O(O)$_{2}$ M\textbf{\textit{O}}enes with a thermal stability (Fig. S17c).

We then study the electronic band structures of 2H-Zr$_{2}$O(O)$_{2}$. Note that the exact gap value has little influence on the main features of the valley characters, so we will base our discussion mainly on the PBE band structures \cite{li2020valley,li2021correlation,yan2023triggering}. Pbands in the absence of SOC is shown in Fig. 7a. Apparently, 2H-Zr$_{2}$O(O)$_{2}$ has a direct band gap of 0.46 eV, together with VBM and CBM located at the same high symmetry K/K' point, which are the two inequivalent corners of the hexagonal Brillouin zone. Moreover, 2H-Zr$_{2}$O(O)$_{2}$ has two valleys at the K and K', which are related to each other by time reversal symmetry. When the relativistic effect is included, the spin projected band structure in Fig. 7b shows that the spin-up and spin-down states near CBM have a splitting of $\sim$85 meV ($\Delta_{C}$), whereas the valley spin splitting (VSS) near VBM  ($\Delta$$_{V}$) is very small ($\sim$1.4 meV). In contrast to group IV and VI TMDs, where the large VSS is generated at VBM. As we known, the large VSS observed at CBM for 2H-Zr$_{2}$O(O)$_{2}$ is extremely rare, reported only in 2H-Tl$_{2}$O ($\sim$610 meV) \cite{ma2018conduction} and 2H-HfN$_{2}$($\sim$314 meV) \cite{mohanta2020coupled}. In subsequent, the orbital contributions give a deeper insight into the different behaviour of VSS at CBM and VBM. The CBM is predominantly contributed by Zr-$\textit{d}_{x^{2}-y^{2},xy}$, and the dominant orbital occupations for VBM are Zr-$\textit{d}_{z^{2}}$. Thus, given that the $\textit{C}_{s}$ symmetry guarantees the out-of-plane potential gradient symmetry, the VSS in 2H-Zr$_{2}$O(O)$_{2}$ arises mainly from the in-plane potential gradient asymmetries. Due to its out-of-plane orientation, Zr-$\textit{d}_{z^{2}}$ has no effect on the VSS at VBM. On the other hand, the in-plane characters of Zr-$\textit{d}_{x^{2}-y^{2},xy}$ states contribute the VSS at CB. Subsequently, the weak SOC strength within the Zr atom results in the tiny splitting for 2H-Zr$_{2}$O(O)$_{2}$, comparable to that of MoSSe ($\Delta_{C/V}$ = 13.7/170 meV) \cite{din2019rashba} and MoSe$_{2}$ ($\Delta_{V}$ = 180 meV) \cite{yang2017considering}. In addition, by choosing a suitable substrate \cite{mohanta2020coupled}, larger VSS can be expected in 2H-Zr$_{2}$O(O)$_{2}$. Similar to 2H-Tl$_{2}$O \cite{ma2018conduction}, the magnitudes of the VSS in both CBM ($\Delta_{C}$) and VBM ($\Delta_{V}$) are robust against mechanical deformation (Fig. S19), favoring its realistic applications.

For two inequivalent K and K' valleys, time-reversal symmetry generally leads to the opposite ordering of spin-up and spin-down states. In this regard, spin can be selectively excited through the optical selection rule. Furthermore, by using various circular polarizations and frequencies of optical illumination, the carriers can be  selectively stimulate with different combinations of spin and valley indexes \cite{yao2008valley,ma2018conduction}. As described in Fig. 7c, the interband transitions at K valley only relates to right-handed circularly polarized light ($\sigma$$^{+}$), and left-handed circularly polarized light ($\sigma$$^{-}$) excites the carriers at  K' valley. Therefore, under the excitations with lef-hand polarized light with a frequency $\omega$$_{u}$, spin-up electrons and spin-down holes can be populated in the K' valley. As for  K valley, spin-down electrons and spin-up holes can be generated when applying right-band polarized light with a frequency $\omega$$_{u}$. Furthermore, the linearly polarized light is the combination of right-handed and left-banded circularly polarized light, and thus can excite the electrons and holes in both K and K' valley. As illustrated in Fig. 7d, when the linearly polarized light is under a frequency $\omega$$_{u}$, spin-up electrons and spin-down holes are regulated to K' valley, while spin-down electrons and spin-up holes can be generated in K valley. In addition, due to the opposite Berry curvatures in CBM and VBM, the excited electrons and holes in the same valley will acquire opposite transverse velocities under an in-plane electric field \cite{ma2018conduction}. This leads to spin-up electrons (holes) from valley K' (K) are accumulated at one boundary, and the spin-down holes (electrons) from valley K' (K) are accumulated at the opposite boundary. In this process, the spin and valley Hall currents occur and  bring the coexistence of spin and valley Hall effects. Since both electrons and holes are accumulated at boundary, the charge neutrality will be maintained, thus, the charge Hall current will not be observed. In this case, each boundary will carry a net spin and a valley polarization. Moreover, at a given boundary, the e-h recombinations are not allowed, because both flip spin and valley indices are required, giving a protected valley-spin coupling.

\section{CONCLUSIONS}
In conclusion, we have systematically investigated the M$_{n+1}$O$_{n}$T$_{2}$ M\textbf{\textit{O}}enes family, and have built an online material library. In addition, the roles of "M" and "\textbf{\textit{O}}" sites are investigated, and 1T-Ti$_{2}$OF$_{2}$ exhibits an ideal two-dimensional Dirac nodal loop state.  What's more surprising is that there are 14 direct semiconductors among the M$_{n+1}$O$_{n}$T$_{2}$ M\textbf{\textit{O}}enes, greatly expanding the direct semiconductors in the MXenes family and extending their light absorption range. In particular, we  focus on several direct semiconductors with unique features in M\textbf{\textit{O}}enes family.  1T-Ti$_{2}$\textit{\textbf{O}}F$_{2}$ (\textbf{\textit{O}}=O, S, Se) have a direct band gap of 0.67$-$1.18 eV (HSE06), and show strong light absorption ability from the infrared to the ultraviolet region. Besides, inspired by the spatially distinguished VBM and CBM states in 2H- and 1T-Y$_{2}$TeO$_{2}$, they possess a long carrier lifetime of 2.38 and 1.24 ns, respectively. In addition, conduction-band VSS is found in 2H-Zr$_{2}$O(O)$_{2}$, providing an alternate candidate for manipulating the coupled spin and valley physics. Of course, the properties of emerging M\textbf{\textit{O}}enes are not limited to these and deserve to be explored in the future.

\section*{ACKNOWLEDGEMENTS}
This work is supported by the Natural Science Foundation of China (Grant No. 12374057), the Startup funds of Outstanding Talents of UESTC (A1098531023601205), National Youth Talents Plan of China (G05QNQR049).  B.-T.W. acknowledge financial support from the Natural Science Foundation of China (Grants No. 11675195 and No. 12074381) and Guangdong Basic and Applied Basic Research Foundation (Grant No. 2021A1515110587).

\bibliographystyle{apsrev4-1}
\bibliography{apssamp}
\end{document}